\documentclass[twocolumn,epsfig,prl]{revtex4}

\usepackage{graphics}
\usepackage{graphicx}
\usepackage{epstopdf}
\usepackage{amssymb}
\usepackage{amsmath}
\usepackage{hyperref}
\usepackage{latexsym}

\begin{document}

\renewcommand{\thefootnote}{\fnsymbol{footnote}} 
\renewcommand{\theequation}{\arabic{section}.\arabic{equation}}

\title{Pair interaction ordering in fluids with random interactions}

\author{Lenin S. Shagolsem$^1$}
\email{lenin.shagolsem@biu.ac.il}
\author{Dino Osmanovi\'c$^1$}
\email{d.osmanovic@ucl.ac.uk}
\author{Orit Peleg$^2$}
\email{opeleg@seas.harvard.edu}
\author{Yitzhak Rabin$^1$}
\email{yitzhak.rabin@biu.ac.il}
\affiliation{$^1$Department of Physics, and Institute of Nanotechnology and Advanced Materials, 
Bar-Ilan University, Ramat Gan 52900, Israel}
\affiliation{$^2$School of Engineering and Applied Sciences, 
Harvard University, Cambridge MA 02138, USA}    


\begin{abstract}
\noindent 
We use molecular dynamics simulations in 2d to study multi-component fluid in the limiting case 
where {\it all the particles are different} (APD). The particles are assumed to interact via 
Lennard-Jones (LJ) potentials, with identical size parameters  but their pair interaction parameters 
are generated at random from a uniform or from a peaked distribution. We analyze both the global 
and the local properties of these systems at temperatures above the freezing transition and find 
that APD fluids relax into a non-random state characterized by clustering of particles according 
to the values of their pair interaction parameters (particle-identity ordering).
\end{abstract}

\maketitle 


Multi-component systems are often encountered in studies of 
colloidal fluids and dispersions, which are polydisperse in particle size. They are 
also quite common in biology:  a cell contains many thousands of different types of proteins
and other macromolecules that differ in their size, shape and in their interactions.  In this work we focus on 
the multiplicity of interactions between the constituents
characteristic of biological systems (we make no attempt to model real proteins) and introduce a minimal physical model in 
which {\it all particles are different} (APD) in the sense that their interaction parameters are 
chosen at random from a given distribution. This model differs in several important ways
from past theoretical studies of multi-component systems that addressed size polydispersity of colloids and  
used a coarse-grained description (in terms of moments of the distribution of particle sizes)
of the thermodynamic properties such as free energy and 
phase diagrams of these systems (see e.g., \cite {Stell, Sollich, Warren}). Conversely, in our APD 
model all particles have identical size parameters and differ only in their attractive interactions with other
particles. We use computer simulations to obtain insights about the temperature-dependent organization of APD fluids into non-random states
in which the pair interaction parameters (PIP) of neighboring particles become strongly correlated. 
Because of the huge configurational space of APD systems (there are $N!$ non-equivalent particle permutations for each spatial configuration of $N$ particles), we prefered to avoid apriori assumptions about equilibration and used molecular dynamics (MD) rather than equilibrium Monte-Carlo methods.

The MD simulations were carried out in 2D under NVT ensemble ($N=2500$ particles, density 
$\rho^\ast=N/{\rm Area}=0.6944$) using the open source package LAMMPS \cite{lammps}. 
All the particles have the same mass $m$ and size $\sigma$ which are set to 
unity, and the simulation box is periodic in both X- and Y- axes. 
The particles interact via Lennard-Jones (LJ) potential 
\begin{equation}
U_{\tiny{_\text{LJ}}}(r) = 4\epsilon_{ij}\left[(\sigma/r)^{12}-(\sigma/r)^{6}\right]~,
\label{eqn: LJ-potential}
\end{equation}
which is cut and shifted to zero at a distance $r_{c}=2.5\sigma$ ($U_{\tiny{_\text{LJ}}}(r>r_{c})=0$). 
The pair interaction parameter $\epsilon_{ij}$ gives the depth of the potential well and $r$ is the separation between a pair of 
particles $i$ and $j$. All the physical quantities are expressed in LJ reduced units \cite{allen}. 
The Langevin equations of motion 
of the particles are integrated with a timestep of $\delta t=0.005\tau_{_{\rm LJ}}$, where
$\tau_{_{\rm LJ}}=\sigma(m/\epsilon)^{1/2}=1$ is the LJ time (we take $\sigma=m=1$ and
define $\epsilon=1$ as the lower bound on $\epsilon_{ij}$). 
The friction coefficient is fixed at $\zeta=1/\tau_d=1/50\tau_{_{\rm LJ}}$ ($\tau_d$ is the viscous 
damping time). We equilibrate the system at high temperature and then bring the system down to the 
required temperature and let it relax for about $5\times 10^4 \tau_{_{\rm LJ}}$  to steady-state in 
which the ensemble-averaged (potential) energy remains constant in time. All
the data reported in this work were obtained in the stationary regime. 

\begin{figure}[ht]
\centering
\includegraphics*[width=0.45\textwidth]{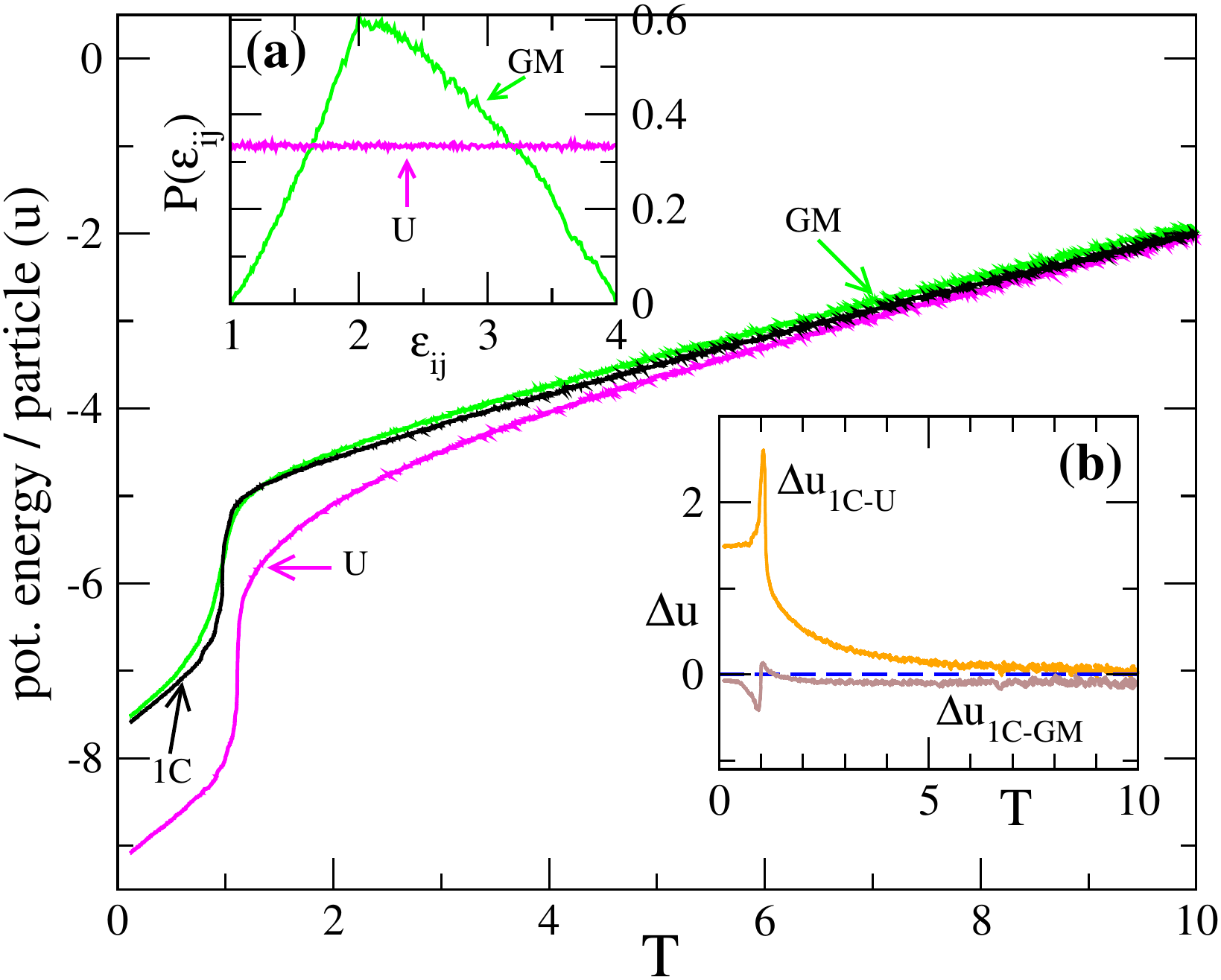}
\begin{center}
\caption{Heating curves of the potential energy per particle for the GM, 
U and 1C (one-component) systems. 
The distribution of PIPs in the range $1-4$ for GM and U 
systems is shown in inset (a), and the energy difference $\Delta u$ between the 1C-system 
and the APD-systems are shown in inset (b).} 
\label{fig: fig1}
\end{center}
\end{figure}

Two different distributions $P(\epsilon_{ij})$ are considered: 
(1) {\it Geometric mean (GM) distribution}: a particle $i$ is assigned an interaction strength 
$\epsilon_i$ in the range $1-4$ taken randomly from a uniform distribution. We then adopt the Berthelot mixing ansatz \cite {berthelot} and define 
$\epsilon_{ij}=\sqrt{\epsilon_i\epsilon_j}$. The distribution $P(\epsilon_{ij})$ 
can be computed analytically using the methods of ref.~\cite{glen} and is shown in inset (a) in fig.~\ref{fig: fig1}. It is peaked at
$\epsilon_{ij}^{\rm mp}=2.0$ and has a mean of $\langle\epsilon_{ij}\rangle=2.42$.\\
\noindent
(2) {\it Uniform (U) distribution}: Instead of assigning each particle a value $\epsilon_i$, random $\epsilon_{ij}$ 
values are assigned to all the possible pairs of particles, i.e., $N(N-1)/2$ 
numbers are drawn randomly from a uniform distribution in the range $1-4$ (horizontal line in inset (a) in fig.~\ref{fig: fig1}). 
Unlike GM where the identity of a particle is determined by its value of $\epsilon_i$, 
in the U system a particle $i$ is characterized by the set of  $N-1$ PIPs
with all other particles ($j\ne i$) and since the probability of generating $2$ identical 
such sets is vanishingly small, we conclude that all particles in the system are different from 
each other (note that since their interaction parameters are taken from a uniform distribution, they 
are all identical in a statistical sense). As a reference we use a one-component (1C) system with 
$\epsilon_{ij}=2.5$ (the average value in the interval $1-4$) for all the pairs. 

In fig.~\ref{fig: fig1} we present the heating curves for the average potential energy per particle 
as a function of $T$ for the two systems, GM and U (we used a heating-cooling rate of $dT/dt\approx10^{-5} 1/\tau_{_{\rm LJ}}$ at which both systems exhibit weak hysteresis upon heating 
and cooling- not shown). The abrupt change of  potential energy with temperature is a signature of a transition from a  liquid to a  solid phase, characterized by the appearance of  positional and orientational (hexatic \cite{nelson}) ordering. A detailed study of the low-temperature solid phase is beyond the scope of this work but preliminary results indicate that the particles form a hexagonal crystal (this crystal may be metastable with respect to PIP ordering). This concurs with the well-known observation that in order to prevent crystallization (which always occurs in simulations of 1C LJ systems), one has to resort to mixtures of particles of different sizes \cite{Kob} (recall that our particles have identical size). The precise nature of the liquid-solid transition in 2D is still controversial even in 1C LJ systems and may be strongly affected by finite size effects \cite{patashinsky}. In this work {\it we concentrate on the statistical properties of APD fluids} and, therefore, we focus on the temperature range $T\ge T^\ast$ where
$T^\ast$ is defined as the temperature at which the radial distribution function decays over half the 
system size. This yields $T^\ast\approx1.0$  and $1.1$, for the GM and the U systems
respectively, the former being close to that of the 1C system. We verified that similar results are obtained if we identify the transition temperature with the peak of the specific heat, with the drop in mean inter-particle distance or with the increase in orientational ordering (not shown). Since the difference between 
the systems stems from different distributions of PIPs, we expect the energy vs. temperature curves to approach each other in the large T limit ($\epsilon_{ij}\ll k_BT$) where 
short-range ($r<\sigma$) repulsion dominates,  and to diverge from each other 
at lower temperatures ($\epsilon_{ij}/ k_BT\sim O(1)$) where the non-universal attractive part of the potential energy plays an increasingly important role; both effects are observed in fig.~\ref{fig: fig1}. 
Inspection of inset (b) in fig.~\ref{fig: fig1} shows that the GM system has a slightly higher energy than the 1C system, possibly because the mean of its 
$P(\epsilon_{ij})$ distribution (2.42) is slightly lower than that of 1C system (2.5).  The larger slope of the potential energy vs. temperature curve for the U system means that the specific heat $C_V \propto\left(\frac{\partial U}{\partial T}\right)_V$ and the entropy change
$\Delta S = \int_{T_0}^{T} (C_V/T) dT$ 
are both higher for the U system. If the energies of the U and the 1C systems were comparable, this would suggest that the liquid-solid transition should take place at lower temperature in the former system, in direct contradiction with our simulation results. This, however, is not the case and the upward shift of the transition temperature ($1.1$ for U vs. $0.97$ for 1C) is the consequence of the large downward shift of the energy of the U system compared to the 1C observed in fig.~\ref{fig: fig1}. As will be shown in the following, this shift reflects the PIP ordering of the U fluid due to the existence of a large number of states in which pairs of particles with high values of $\epsilon_{ij}$ cluster together, thus reducing the mean energy of the system. Although GM fluids undergo  PIP ordering as well, its extent is more limited because of the constraints imposed by the peaked form of the distribution function of the pair interaction parameters (see inset (a) in  fig.~\ref{fig: fig1}). 

We computed the probability distribution of the total energy and found that at high temperatures the distributions for all three systems are strictly Gaussian with width obeying the thermodynamic relation between the variance of energy fluctuations and the specific heat \cite{landau}. As the transition temperature is approached, deviations from Gaussian behavior are observed as the correlation length approaches the finite system size \cite{joubaud2008}, but while low energy fluctuations are enhanced and high energy ones are suppressed for the U fluids, the reverse takes place for GM and 1C fluids (see SI for plots of global energy distributions). This is another manifestation of higher  PIP ordering in the U system.

\begin{figure}[ht]
\centering
\includegraphics*[width=0.45\textwidth]{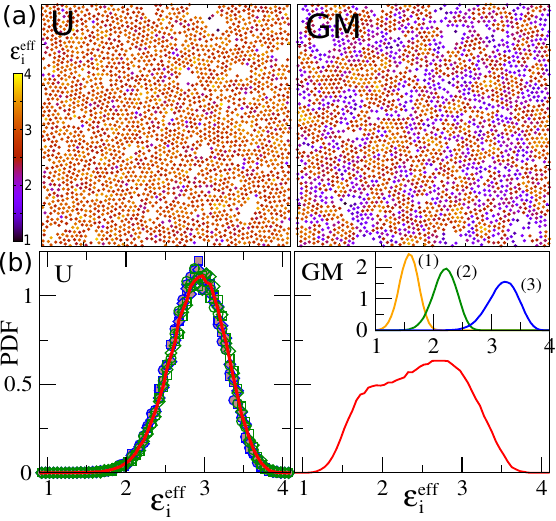}
\begin{center}
\caption{(a) Typical configuration of U and GM systems at $T=1.2$ and $T=1.1$ 
respectively, that correspond to $\delta=(T-T^\ast)/T^\ast\approx 0.1$. 
The particles are colored according to $\epsilon_i^{\rm eff}$ values. 
(b) Probability density functions (PDF) of $\epsilon_i^{\rm eff}$ for the APD 
systems. U system: Symbols are for the data obtained from tracking 
of three randomly chosen particles. Open and filled symbols represent two independent 
realizations of $\epsilon_{ij}$ values. The red curve represents the ensemble average. 
GM system: Ensemble average data is shown. The inset shows the PDFs for 
particles with $\epsilon_i$ values in the range (1) 1-1.2, (2) 2.9-3.1 and 
(3) 3.8-4.}
\label{fig: fig2}
\end{center}
\end{figure}

We proceed to take a closer look into the PIP ordering of APD fluids.
In the GM fluid there is hierarchy in the spatial organization of particles 
according to $\epsilon_i$ values, i.e., particles with larger values of $\epsilon_i$ tend 
to form clusters at intermediate $T$ (around $T^\ast$) while particles with small $\epsilon_i$ 
are preferentially localized around vacancies in the dense phase (see fig.~\ref{fig: fig2}(a)). 
This suggests that GM fluids relax to a locally-ordered state in which there are strong 
correlations between the interaction parameters of neighboring particles. In order to characterize 
the local statistical properties of both GM and U fluids (in the U system there is no single parameter $\epsilon_i$ that characterizes the $i^{\rm th}$ particle) 
we introduce the {\it effective interaction parameter $\epsilon_{i}^{\rm eff}$} of  particle $i$, 
that depends not only on the intrinsic interaction parameters of the particle, but also on the identity of  its 
nearest neighbors in a particular configuration of the system,
\begin{equation}
\epsilon_{i}^{\rm eff}=\sum_{j=1}^{n_b}\epsilon_{ij}/n_b,
\end{equation}
where the sum over $j$ goes over all the $n_b$ surrounding particles within a cut-off radius 
$r_c=1.7$ that corresponds to the minimum between the first and the second peak of the radial 
distribution function. In fig.\ref{fig: fig2}(a) we show typical snapshots of the two APD systems taken 
at the same normalized distance $\delta=(T-T^\ast)/T^\ast\approx 0.1$ from their respective transition 
temperatures (the particles are colored according to the values of $\epsilon_{i}^{\rm eff}$) (see SI for movies). 
Inspection of this figure shows that the U system is  more homogeneous and has a higher mean value 
of $\epsilon_{i}^{\rm eff}$ than the GM system (for the 1C system the distribution is a delta function at $\epsilon_{i}^{\rm eff}=2.5$). This observation is confirmed quantitatively in 
figs.\ref{fig: fig2}(b) where we plotted the distributions of $\epsilon_{i}^{\rm eff}$ for both systems. While the U distribution 
is nearly Gaussian and is centered about  $\epsilon_{i}^{\rm eff}\simeq 2.9$, the GM distribution is much broader, 
with a complex shape that results from the constraints imposed by the non-uniform distribution 
$P(\epsilon_{ij})$. Further insight into the PIP ordering 
of the GM fluid can be gained by considering particles with $\epsilon_i$  in a particular narrow 
interval of values and computing the distributions of $\epsilon_{i}^{\rm eff}$ for three different 
such intervals. We find (see inset fig.\ref{fig: fig2}(b)) that the corresponding distributions are nearly Gaussian, with width increasing with $\epsilon_i$.  Since each of the 
peaks is much narrower than the cumulative distribution, the GM fluid behaves as a 
mixture of finite number ($\ll N$) of components that undergoes microphase separation as $T^*$ is approached 
(see fig.\ref{fig: fig2}(a)). 

\begin{figure}[ht]
\centering
\includegraphics*[width=0.49\textwidth]{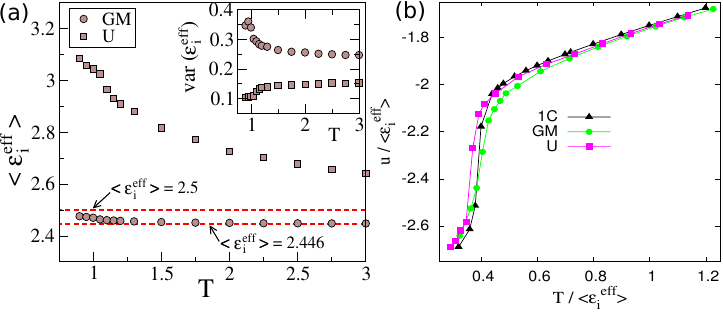}
\begin{center}
\caption{(a) $T$-dependence of the mean and variance of the distributions of $\epsilon_i^{\rm eff}$ 
for the two APD systems. (b) The potential energy per particle as a function of the temperature, both scaled by $\langle\epsilon_{i}^{\rm eff}\rangle(T)$.} 
\label{fig: fig3}
\end{center}
\end{figure}

We now turn to study the temperature dependence of the distributions 
of $\epsilon_{i}^{\rm eff}$. Inspection of fig.\ref{fig: fig3}(a) shows that
at high temperatures, the means of the U and the GM distributions of $\epsilon_{i}^{\rm eff}$ approach the means of the corresponding $\epsilon_{ij}$  distributions, $2.5$ and $2.42$, respectively. As temperature is decreased the two distributions shift to higher values of $\epsilon_{i}^{\rm eff}$ and, while the width of the U distribution decreases, that of the GM distribution increases. The upward shift of $\langle\epsilon_{i}^{\rm eff}\rangle$  is much more pronounced for the U than for the GM system, a signature of more efficient PIP ordering in the former system. Preliminary results of our Monte-Carlo simulations of the crystaline phase of the U system indicate that as temperature is lowered below $T^*$,  the peak of the distribution continues to shift to higher values of $\epsilon_{i}^{\rm eff}$ and the distribution becomes asymmetric with a tail towards lower values of $\epsilon_{i}^{\rm eff}$ (not shown). We believe that at yet lower temperatures the distribution narrows down and approaches a $\delta$-function as $T\rightarrow 0$ (in the limit $N\rightarrow\infty$).
Since $\langle\epsilon_{i}^{\rm eff}\rangle$ can be thought of as new temperature-dependent energy scale, it is interesting to check whether the thermodynamics of the two multi-component systems can be reduced to that of an effective one-component system \cite{Rice} by rescaling all energies and temperatures by $\langle\epsilon_{i}^{\rm eff}\rangle(T)$. As shown in fig. \ref{fig: fig3}(b) this anzatz appears to work for both the U and the GM systems (in the latter case, some qualitative differences with the 1C system remain upon rescaling).

\begin{figure}[ht]
\centering
\includegraphics*[width=0.45\textwidth]{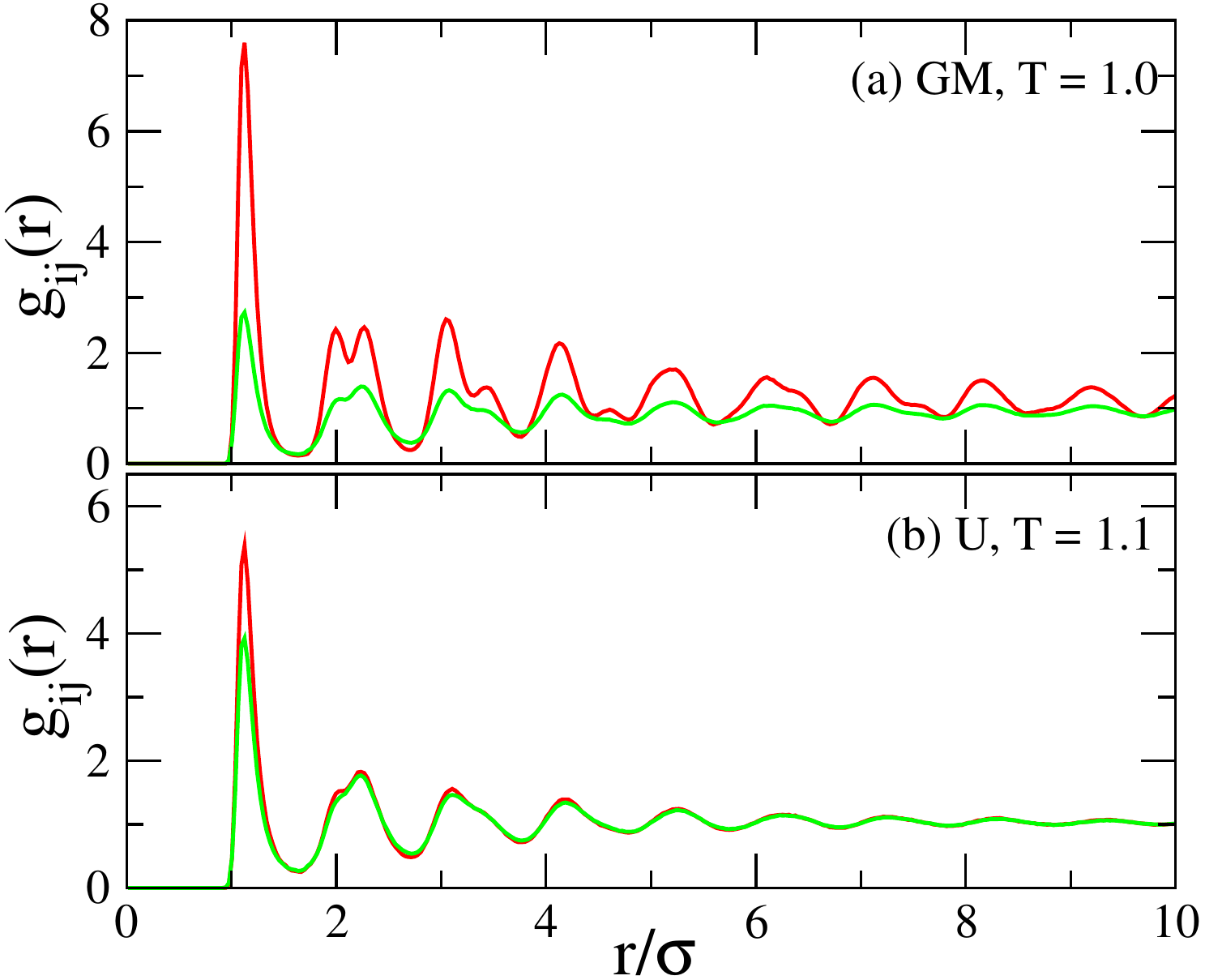}
\begin{center}
\caption{Partial pair correlation function $g_{_{\rm ij}}(r)$ of GM and U systems (at their respective $T^\ast$) 
obtained by grouping particles according to their $\epsilon_{i}^{\rm eff}$ values (see text). 
Diagonal and off-diagonal PRDFs are shown in red and green colors, respectively.}
\label{fig: fig4}
\end{center}
\end{figure}

In order to gain additional insight into the nature of self-organization of APD fluids, we calculated the partial radial distribution function (PRDF) 
$g_{_{\rm ij}}(r)$ that gives the probability to find a particle of type $j$ at a distance $r$ from particle $i$, for GM and U fluids at their respective transition temperatures (see fig. \ref {fig: fig4}). Here $i$ stands for particles with $\epsilon_{i}^{\rm eff} > \langle\epsilon_{i}^{\rm eff}\rangle$ and $j$ denotes either particles in the same (diagonal PRDF) or in the complementary range $\epsilon_{j}^{\rm eff} < \langle\epsilon_{j}^{\rm eff}\rangle$ (off-diagonal PRDF). Here $\langle\epsilon_{i}^{\rm eff}\rangle \approx 2.47$ (GM) and $2.97$ (U).  In the U system the only difference between the diagonal and the off-diagonal PRDFs is that the first peak is slightly higher for the former, in accord with our assertion that the PIP distribution in U fluids is spatially homogeneous. In the GM system, all the diagonal peaks are higher than the off-diagonal ones, indicating that GM fluids relax into an inhomogeneous state in which regions enriched in particles with larger values of $\epsilon_{i}$ possess higher long-range (hexagonal) order.

We would like to stress that the distributions in fig.\ref{fig: fig2}(b), as well as all other statistical properties we considered so far, 
do not depend on the particular realization of the chosen $\epsilon_{ij}$ distribution, i.e., they are translationally ergodic \cite {stein and newman}. 
In order to check whether APD fluids are ergodic in the usual sense of the word, we calculated the distribution of $\epsilon_{i}^{\rm eff}$ by (a) sampling the instantaneous configuration of the system (ensemble average) and averaging over many instantaneous configurations in order to reduce the noise, and by (b) following the trajectories of several labeled particles over large time intervals and collecting the values of $\epsilon_{i}^{\rm eff}$ along these trajectories. As shown in fig.\ref{fig: fig2}(b) both methods yield identical results for the U system (in the GM system trajectory sampling will yield different results for particles with different values of $\epsilon_{i}$).

To summarize, we have studied the steady-state properties of systems of particles with random PIPs. We introduced an effective interaction parameter $\epsilon_{i}^{\rm eff}$ that  characterizes the degree of local PIP ordering and found that this parameter increases monotonically upon cooling as the freezing transition is approached (preliminary results of Monte-Carlo simulations of 2D APD crystals suggest that the increase of $\epsilon_{i}^{\rm eff}$ saturates upon further lowering of temperature).
The constraints that define this organized state confine the system to a subspace of the available 
configurational space but the dimensionality of this subspace appears to be sufficiently large to 
allow for its efficient exploration (for discussion of diffusion in rugged high-dimensional energy and fitness landscapes see 
ref.\cite{rugged}) and therefore, the macroscopic behavior of APD fluids is similar to that 
of normal fluids . The degree of PIP ordering depends on the distribution of PIPs and is much more pronounced in U than in GM systems. 
This is the consequence of the fact that the choice of PIPs is much more constrained in the GM than in the U system (we choose $N$ random numbers for GM vs. $N^2$ random numbers for U), and due to the unbiased (flat) vs. peaked forms of the PIP distributions of U and GM systems, respectively.

One may wonder whether APD systems obey standard thermodynamics 
since the fact that all particles are distinguishable suggests that there is a non-extensive contribution to the entropy ($\ln N!$)
associated with all possible permutations of the particles. However, in accord with general arguments 
about the resolution of the Gibbs paradox in classical systems of distinguishable colloidal particles \cite{Warren,frenkel}, 
we did not observe any violations of standard thermodynamic relations.
Results on the dynamics of the systems, e.g., singe particle tracking (SPT) measurements which is 
more relevant to make connection with SPT experiments on complex (biological) systems 
(see, e.g., \cite {izzedin, burov}) will be discussed in an upcoming publication.
Other distributions of pair interaction parameters and 3D APD systems will be studied in future work.

 \begin{acknowledgements}
Helpful discussions with E. Braun, D. Rapaport, A. Ermolin, D. Kessler and E. Barkai are greatfully acknowledged. We would to thank I. Parshani, R. Shalev and M. Caspi for help with the simulations. YR's research was supported by the I-CORE Program of the Planning and Budgeting committee and the Israel Science Foundation, and by the US-Israel Binational Science Foundation. 
\end{acknowledgements}

\newpage
\onecolumngrid
\renewcommand{\thepage}{S\arabic{page}}  
\renewcommand{\thesection}{S\arabic{section}}   
\renewcommand{\thetable}{S\arabic{table}}   
\renewcommand{\thefigure}{S\arabic{figure}}
\renewcommand{\theequation}{S\arabic{equation}}
\setcounter{figure}{0}
\setcounter{equation}{0}
\setcounter{table}{0}
\newpage
\appendix
\begin{center}
{\Large SUPPLEMENTARY MATERIAL}
\end{center}


\section{Global energy fluctuations}

\begin{figure}[ht]
\centering
\includegraphics*[width=0.5\textwidth]{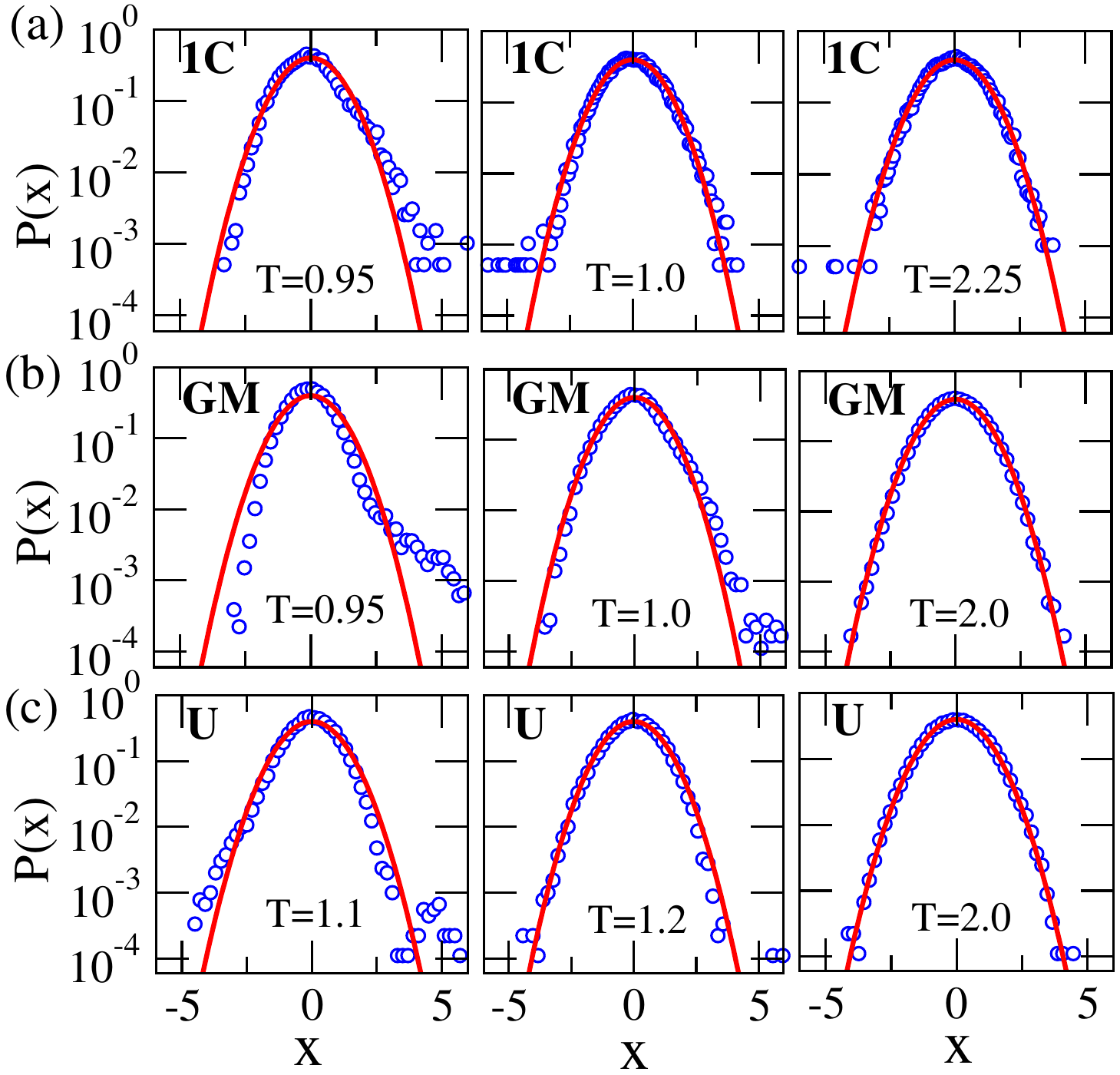}
\begin{center}
\caption{PDF of $x$ for the 1C and APD systems at three different 
temperatures indicated in the figure. The solid lines are the Gaussian PDF with zero-mean and 
unit variance. For temperatures approaching $T^\ast$ the PDF deviates from a Gaussian PDF.} 
\label{fig: sfig1}
\end{center}
\end{figure}

\noindent The distributions of the potential energy per particle $u=U/N$ 
(with $U$ the total potential energy, and $N$ total number of particles) are 
obtained for temperatures above and close to $T^\ast$. We rescale $u$ to to get a distribution 
with zero mean and unit variance, i.e., we define a variable $x\equiv(u-\langle u\rangle)/\sigma_u$, 
with $\langle u\rangle$ the ensemble average and $\sigma_u^2$ the variance and obtain 
the probability density function $P(x)$ shown in fig.\ref{fig: sfig1}.

\end{document}